\documentclass[conference]{IEEEtran}
\pdfoutput=1
\usepackage{blindtext, graphicx}
\usepackage{epstopdf}
\usepackage{tabularx}
\usepackage{amsmath}
\usepackage{gensymb}
\begin{document}

\title{  Optimization of LTE Radio Resource Block Allocation for  Maritime Channels}

\author{\IEEEauthorblockN{Amit Kachroo\IEEEauthorrefmark{1},
Mehmet Kemal Ozdemir\IEEEauthorrefmark{2},
Hatice Tekiner-Mogulkoc\IEEEauthorrefmark{3}} \\
\IEEEauthorblockA{\IEEEauthorrefmark{1}\IEEEauthorrefmark{2}\IEEEauthorrefmark{3}Graduate School of Natural and Applied Sciences\\
Istanbul Sehir University,
Istanbul, Turkey\\ Email:amitkachroo@std.sehir.edu.tr\\kemalozdemir,haticetekiner\{@sehir.edu.tr\}} }

\maketitle

\begin{abstract}
In this study, we describe the behavior of LTE over the sea  and  investigate the problem of radio resource block allocation in such SINR limited  maritime channels. For simulations of such sea environment, we considered a network scenario of Bosphorus  Strait  in Istanbul, Turkey with different number of ships ferrying between two ports at a given time. After exploiting the network characteristics, we formulated and solved the radio resource allocation problem by  max-min integer linear programming  method.  The radio resource allocation fairness in terms of Jain's fairness index  was computed and it was compared with round robin and opportunistic methods. Results show that  the max-min optimization method performs better than the opportunistic and round robin methods. This result in turn reflects that the max-min optimization method gives us the  high minimum best throughput as compared to other two methods considering  different ship density scenarios in the sea. Also, it was observed that as the number of ships begin to increase in the sea, the max-min  method performs significantly better with good fairness as compared to the other two  methods.
\end{abstract}


\section{Introduction}
 
\begin{figure}[!ht]
\centering
\includegraphics[width=8.5cm]{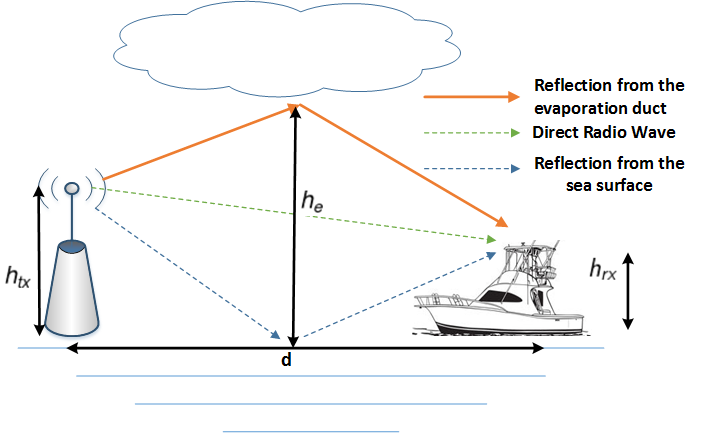}
 \caption{ 3-Ray path loss model, where $h_{tx}$ is the height of the transmitter, $h_{rx}$ is the height of receiver, $h_{e}$ is the evaporation duct height and $d$ is the distance between the transmitter and receiver.}
\label{evaporation}
\end{figure}

 There are extensive number of studies in the analysis of LTE performance  for urban landscape \cite{pathloss1}-\nocite{pathloss2}\cite{pathloss3}.  These models are generally based on Okumara-Hata Model, COST 231-Hata Model, Ikegami Model, or 2-Ray model \cite{book1,cost231}. Okumara-Hata model considers open, suburban, and urban areas for measuring path loss, while COST 231-Hata Model is just an extended form of Okumara-Hata Model that considers frequencies from 1500 MHz to 2000 MHz as compared to 150 MHz to 1500 MHz in Okumara-Hata model. On the other hand, Ikegami Model  gives deterministic prediction of field strength at specific point but underestimates loss at large distances in urban or suburban areas. In contrast to LTE performance for urban landscape, not much has been done in the field of radio propagation over sea or  LTE  in sea environment. Only few noted research  can be found in the literature \cite{RDS:RDS5428}-\nocite{5089553}\cite{lee2014near}.\\
 \indent From these literature's, the existence of the evaporation duct can be confirmed over the sea all the time. This is the main difference between  the path loss propagation in  sea environment and  path loss propagation in urban environment . In other words, apart from direct line of sight and reflections from sea surface (ground) there is a  reflection from the evaporation duct making it a  multi path loss model that looks like a 3-Ray path loss. Fig.~\ref{evaporation} shows this typical 3-Ray model. \\

 Now, the radio resource allocation problem in LTE networks  has been also  an extensive research topic for long \cite{seong,bazzi}. It begins from a small resource block called Radio resource block (RB) that is assigned to a user \cite{Ghosh:2010:LNW:1840478.1840482}  to cater its service demand that can range from few kilobits per second (Kbps) to some megabits per second~(Mbps).\\
 
 \begin{figure}[!ht]
\centering
\includegraphics[width=8.7cm]{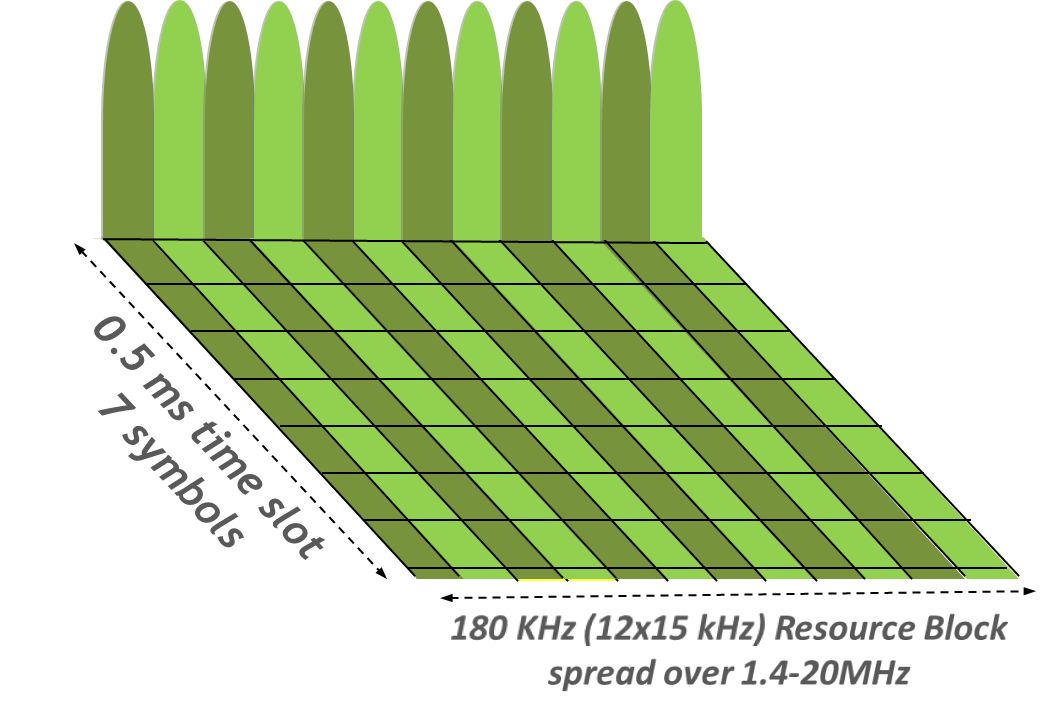}
 \caption{A radio resource block.}
\label{rb}
\end{figure}
 
 A RB has 12 orthogonal frequency division multiplexing (OFDM) subcarriers that are adjacent to each other with a spacing of 15 kHz between two adjacent subcarriers. Each RB (Fig.~\ref{rb}) consists of two sub-time slots of 0.5~ms and each of these sub-time slot utilizes 6 OFDM symbols when normal cyclic prefix is used and  7 OFDM symbols when extended cyclic prefix is used. In RB assignment, the channel state information plays a vital role~\cite{lopez2011optimization} and this  information is acquired by an eNodeB from its connected users periodically. Based on this information, an eNodeB  decides upon the modulation and coding scheme (MCS) \cite{mcs} and the number of radio blocks that it  needs to allocate  to its connected  users. However, in LTE downlink, if a user has been assigned to more than one RB, all these  RBs must  have the same MCS. This increases the complexity of the  radio resource allocation problem.\\
\indent In this study, the max-min optimization technique  that is  used extensively   in wifi optimization \cite{kachroo,pioro}  is leveraged for marine channels. However, some early research on resource block optimization  that used the max-min approach relied on the  characteristics of the urban channels and hence have a different settings than this study \cite{lopez2011optimization,maxmin2}.  To our best knowledge, no studies have been performed for  LTE resource allocation over the sea channels. Hence, this much needed work fills the gap for such case. The rest of the paper is organized as follows: In Section II, we define the LTE-SINR path loss modelling in sea environment. In Section III, we introduce LTE system parameters and the problem formulation. Section IV presents the simulation results and discussions. Lastly, the final conclusion and future work are given in Section V.

  \begin{table}[!ht]
\caption{MCS (Modulation and Coding Schemes), SINR, and MCS Efficiency.}

\centering
\label{table0}
\begin{tabular}{|c|c|c|c|c|}
\hline
\textbf{MCS} & \textbf{Modulation} & \textbf{Code Rate} & \textbf{\begin{tabular}[c]{@{}c@{}}SINR \\ Threshold {[}dB{]}\end{tabular}} & \textbf{\begin{tabular}[c]{@{}c@{}}Efficiency\\ {[}bits/symbol{]}\end{tabular}} \\ \hline

MCS1 & QPSK & 1/12 & -6.5 & 0.15 \\ \hline
MCS2 & QPSK & 1/9 & -4 & 0.23 \\ \hline
MCS3 & QPSK & 1/6 & -2.6 & 0.38 \\ \hline
MCS4 & QPSK & 1/3 & -1 & 0.60 \\ \hline
MCS5 & QPSK & 1/2 & 1 & 0.88 \\ \hline
MCS6 & QPSK & 3/5 & 3 & 1.18 \\ \hline
MCS7 & 16QAM & 1/3 & 6.6 & 1.48 \\ \hline
MCS8 & 16QAM & 1/2 & 10 & 1.91 \\ \hline
MCS9 & 16QAM & 3/5 & 11.4 & 2.41 \\ \hline
MCS10 & 64QAM & 1/2 & 11.8 & 2.73 \\ \hline
MCS11 & 64QAM & 1/2 & 13 & 3.32 \\ \hline
MCS12 & 64QAM & 3/5 & 13.8 & 3.90 \\ \hline
MCS13 & 64QAM & 3/4 & 15.6 & 4.52 \\ \hline
MCS14 & 64QAM & 5/6 & 16.8 & 5.12 \\ \hline
MCS15 & 64QAM & 11/12 & 17.6 & 5.55 \\ \hline
\end{tabular}
\end{table}
 
\section{LTE-SINR path loss Modelling in Sea Environment}

\begin{figure}[!ht]
\centering
\includegraphics[width=8.7cm]{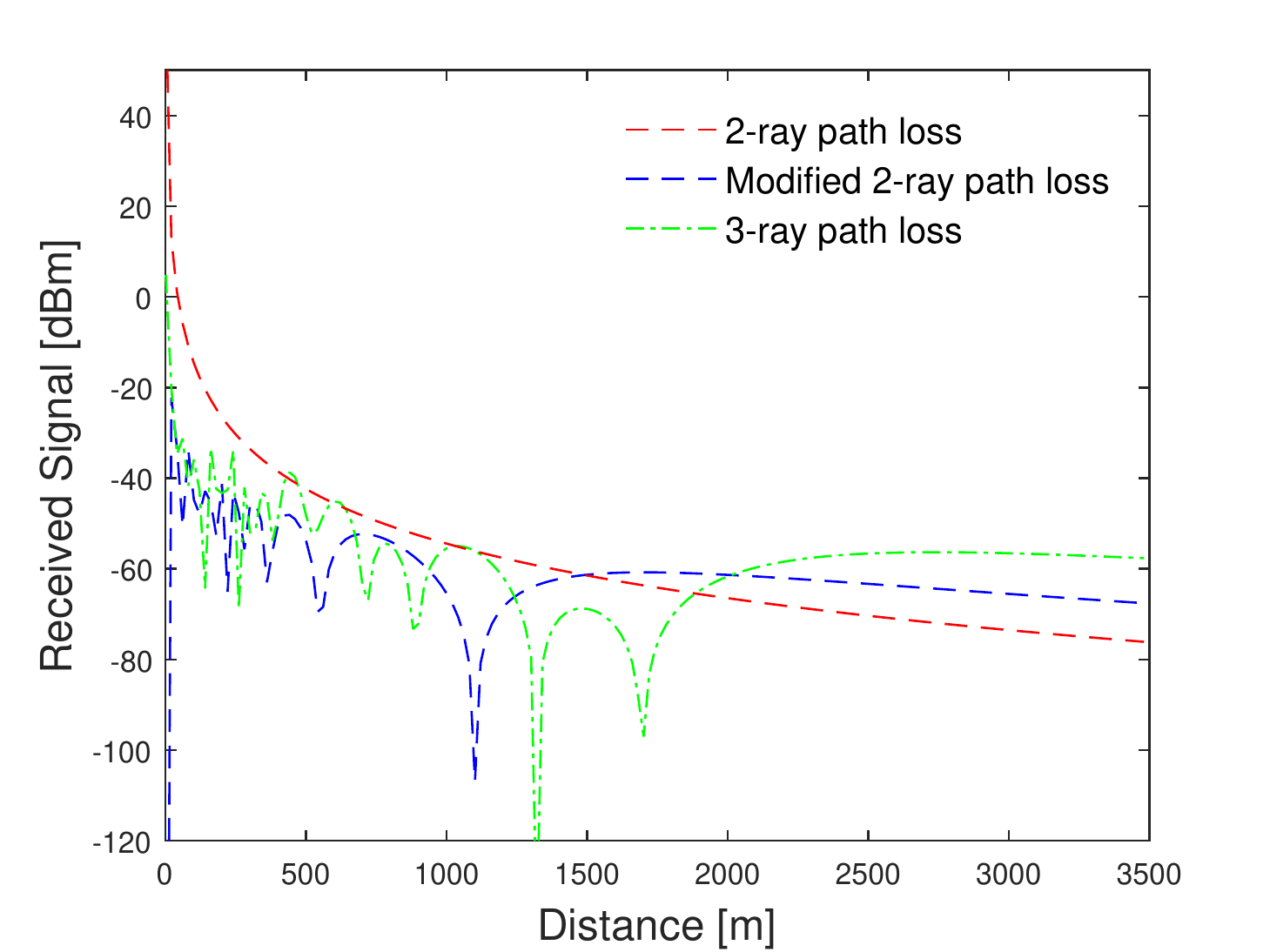}
 \caption{Simulations for different path loss models for sea channels. }
\label{path_loss}
\end{figure}

In this section, we will describe the important consideration for path loss modelling in sea environment. Simulations results as seen in  Fig.~\ref{path_loss} show  that the received signal  over distances in the case of 3-Ray path loss model  is not  as flat when compared to 2-Ray path loss model. It is because of the signals that are  received at the receiver by reflections from the sea surface and the evaporation duct. Although, modified 2-Ray model resembles the 3-Ray model but for LTE over sea, a better estimate  of the received signal power is determined by 3-Ray path loss model as it gives near to practical results \cite{lee2014near}.  Mathematically, the  2-Ray path loss propagation model, 2-Ray modified path loss propagation model, and 3-Ray path loss propagation  model \cite{lee2014near}  can be  represented as:

 \begin{eqnarray}\label{2ray}
  L_{2-Ray}= -10log_{10} \left( \frac{(h_{tx}\cdot h_{rx})^2}{d^4\cdot I } \right ) \nonumber
 \end{eqnarray}
 
 \begin{equation} \label{2Mray}
L_{ 2-RayMod.}=-10log_{10}\left( \left (\frac{\lambda}{4\pi d} \right )^2 \left({2\sin\frac{2\pi h_{tx}h_{rx}}{\lambda d}}\right)^2\right) \nonumber
 \end{equation}
 

\begin{equation} \label{3rayeqn}
L_{3-Ray}=-10log_{10} \left( \left (\frac{\lambda}{4\pi d} \right )^2 \times  \left ( {2(1+\triangle})\right)^2\right) \nonumber
 \end{equation}
\\ \indent with
\begin{eqnarray} \label{delta}
\triangle=2\sin \left (\frac{2\pi h_{tx}h_{rx}}{\lambda d} \right )\times \sin \left (\frac{2\pi (h_{tx}-h_{e})(h_{e}-h_{rx})}{\lambda d} \right)  \nonumber
 \end{eqnarray}
 
  with the parameters being\\ \\
 \indent \indent$\lambda $ \enspace \enspace \enspace: Wavelength in meters\\
   \indent \indent$h_{tx}$ \enspace : Height of transmitter in meters\\
   \indent \indent$h_{rx} $ \enspace : Height of receiver in meters \\
\indent \indent $h_{e}$  \enspace \enspace : Height of evaporation duct \\
\indent \indent $I$ \enspace \enspace \enspace: System loss parameter \\
 \indent \indent $d$ \enspace \enspace \enspace: Distance between transmitting and receiving\\
\indent \hspace{1.5cm} stations.

\begin{figure}[ht]
\centering
\includegraphics[width=8.5cm]{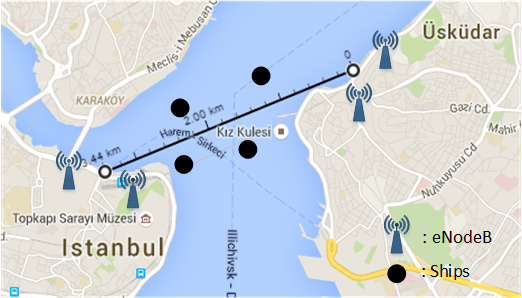}
 \caption{Eminonu and Uskudar ferry ports as seen on Google Maps\textsuperscript{TM} with assumed ships and eNodeB positions. }
\label{google_map}
\end{figure}

 For simplicity, in this study we have assumed $h_{e}$   to be around 25 meters.  As for our network scenario, we selected two ferry ports: Uskudar and Eminonu  in Istanbul, Turkey. Fig.~\ref{google_map} shows these ferry ports on Google Maps\texttrademark \hspace{0.15cm} with assumed ship and eNodeB positions. The distance between these two ports is around 3.7 km and the distance between the two base stations on each port is around 500 meters. At any given moment, there are around 4 to 12 ships travelling from one ferry port to the other. The two lanes "Eminonu to Uskudar" and "Uskudar to Eminonu" are separated by around 300-400 meters. To represent ships, aka users, we choose  equidistant points  in the sea lane,  and to represent eNodeB's we fixed  4 points on the land just near to the ports. The ship distances are assumed uniform in the sea.

\section{LTE system parameters and the problem formulation} 
 
 In this section, we define all the assumptions and the  LTE system parameters that are required to describe the radio resource allocation problem. Problem formulation with different radio resource allocation methods  and their solution are also described in detail in this section.
 
 \subsection{Assumptions}
 The  two basic fundamental assumptions \cite{zabini} regarding our allocation method are : 
a) Throughput perceived by any user $j$ from the connected eNodeB $i$  is a function of how many resource element has been allocated rather than what those resources are,
b) Throughput increases strictly for a user if more resources are allocated to it.\\
\indent Denoting $\beta_{j}$ as a normalised ratio of resource assigned from an eNodeB $i$ to a user $j$ with $\beta_{i}\in [0,1]$, we have

 \begin{equation} \label{beta}
\beta_{j}\overset{\Delta}{=}\frac{r_{ij}}{s_{i}} \nonumber
 \end{equation}
 
where $r_{ij}$ are the resources allocated by an eNodeB $i$ to a user $j$ and $s_{i}$ are the total available resources with the eNodeB $i$. Thus, the  throughput $(T_{j})$ can be written as a strictly increasing function of $\beta$ as:

 \begin{equation} \label{function}
T_{j}= g_{j}(\beta_{j}). \nonumber
 \end{equation}

The assumptions described earlier takes care of orthogonal resource allocation and channel state information availability. Also, it can be intuitively deduced that the throughput would be maximum, if

 \begin{equation} \label{function2}
\sum_{j=0}^{J}\beta_j=\sum_{j=0}^{J}\psi_{j}\left (T_{j}\right)=1  \nonumber
 \end{equation}
 
where
  \begin{equation} \label{function3}
\psi_{j}(T_{j})\overset{\Delta}{=}g_{j}^{-1}.  \nonumber
 \end{equation}
Now, we will define the LTE system parameters.
 \subsection{LTE system parameters:}
 
 The LTE system parameters involved in the problem formulation are described as follows:

      \begin{itemize}
      \item $user\ j=1,2,3,....,J;$  where a user represents a ship in our case.
       \item $eNodeB\ i=1,2,3,....,I;$
       \item $n=1,2,3,....,N;$  where $n$ is the number of resource blocks. 
              
     \item $SINR\ calculation\ parameters$ : For a  user $j$ that is connected to an eNodeB $i$, the SINR can be given by:\\
    \begin{eqnarray}
    SINR_{i,j}=\frac{Power_{i,j}\cdot\Gamma_{i,j}}{\sum_{m=1,m \neq i }^{I}Power_{m,j}\cdot \Gamma_{m,j} + \sigma^2} \nonumber
    \end{eqnarray}
   where $\Gamma_{i,j}$ is the channel gain between eNodeB $i$  and  user $j$, and  $\sigma^2$ is the zero mean noise variance. This $SINR$ represents the link conditions and is used to  determine the connection and throughput between a user and an eNodeB.
    \item $Throughput\ space\ matrix$ :  We represent the throughput space matrix as:
    \begin{equation}
    Th \left [I \right ] \left [J \right ] = \bar{T}_{ij}. \ \forall i \in [1,I],j \in [1,J]  \nonumber
    \end{equation}
    \label{th}
    \noindent The throughput space matrix entries ($\bar{T}_{ij}$'s) is defined as throughput per RB between the user $j$ and eNodeB $i$ and is calculated on the basis of MCS values with respect to SINR levels \cite{lopez2011optimization,mcs} as given in Table~\ref{table0}. 
    The demand per user is represented as :  \\
    \begin{equation}
   D_{j} =\phi \nonumber
    \end{equation}
     where $\phi$ is the minimum best throughput that can be guaranteed to a user in the network.
    \end{itemize} 
Finally, Table~\ref{table} summarizes LTE system  parameters that were used in the simulations. Since we are utilizing 2x2 MIMO, we assume that the data rate is doubled as compared to the data rate of a single  antenna mode. Lastly, for the sake of simplicity the noise variance is taken~as~1.
 
\begin{table}[!ht]
\normalsize
\caption{Simulation Parameters} 
\centering
\begin{tabular}{|l|l|}
\hline
\textbf{Parameters} & \textbf{Values} \\ \hline
Carrier Frequency & 2750 MHz \\ \hline
Number of RBs per eNodeB & 25 \\ \hline
\begin{tabular}[c]{@{}l@{}}OFDM data symbol per sub\\ time slot\end{tabular} & 7 \\ \hline
Antenna & 2x2 MIMO \\ \hline
eNodeB Tx power & 43 dBm \\ \hline
Height of Tx & 20 meter \\ \hline
Height of Rx & 3 meter \\ \hline
Height of Evaporation Duct & 25 meter \\ \hline
Cable loss & 3 dBm \\ \hline
Antenna Pattern & Omnidirectional \\ \hline
Carriers per RB & 12 \\ \hline
Noise Variance  & 1 \\ \hline
\end{tabular}
\label{table}
\end{table}

 \subsection{Problem formulation:}

Given all these assumptions and system parameters, we now describe the problem formulation of  the different radio resource allocation methods.

\subsubsection{Max-min problem formulation}

As a starting point, we assume that the users are  in the transmission range of all eNodeB's and can connect to any one of those eNodeB. Also, it is assumed that all the eNodeB's having perfect knowledge of channel state information and thus, have  the knowledge of throughput space matrix. The  max-min method is formulated in such a way  that it tries to allocates RBs to the worst throughput links first rather than the better throughput links. This in turn  maximizes the minimum throughput of the links in the network and thereby, increases fairness in radio resource allocation. The objective function can be written as Eq. \ref{phi} or can be better simplified in two step as  Eq. \ref{phi0} and   Eq. \ref{objective25}. Also, initially the demand is assumed to be relaxed that is $\phi=0$ .\\

 \begin{eqnarray}
 \max\min_{i \in I}\hspace{0.2cm} \arg \left ( \sum_{i=1}^{I}(\bar{T}_{ij}\times K_{ij}) \right ) \hspace{0.5cm} \forall\ j \in [1,J] \label{phi}\\
  Maximize\ (\phi) \label{phi0} \\
{\sum_{i=1}^{I} \left ( \bar{T}_{ij}\times K_{ij}\right) \ge \phi, \hspace{0.5cm} \forall \ j \in [1,J]} \label{objective25}
\end{eqnarray}

The constraint space is given by:

\begin{eqnarray}
Y_{ij}\in \{0,1\} \label{binary05}\\
\sum_{j=1}^{J}K_{ij} \le N, \hspace{0.5cm} \forall \ i \in [1,I]  \label{binary25} \\
 \sum_{i=1}^{I}Y_{ij} = 1, \hspace{0.5cm} \forall \ j \in [1,J] \label{bin5}\\
  K_{ij} = Y_{ij}\times N, \hspace{0.5cm} \forall \ i,j  \label{relation5}\\
  K_{ij} \geq (D_{j}/ \bar{T}_{ij})\times Y_{ij}. \hspace{0.5cm}  \forall \ i,j \label{capa5}
\end{eqnarray}

Eq.~(\ref{binary05}) defines a binary decision variable $Y_{ij}$ to denote a connection between an eNodeB and a user. Also, an  integer decision variable, $K_{ij}$,  is defined to signify  the number of RBs  that can be allocated from an eNodeB $i$ to a user $j$ in the network. Eq.~(\ref{binary25})  gives a constraint that the  maximum RBs that can be allocated to users from a given eNodeB can be $N$. In Eq.~(\ref{bin5}), $\left ({\sum_{i=1}^{I}Y_{ij}}\right)$ is 1 if an  eNodeB $i$ is connected to a user $j$, else it is zero. On the other hand, Eq.~(\ref{relation5}) makes sure that the RBs ($K_{ij}$)  are allocated to the only  defined connection ($Y_{ij}$). Finally, Eq.~(\ref{capa5}) gives  the capacity constraint with respect to the demand $D_{j}$.\\
\indent In addition to this formulation, Karush-Kuhn-Tucker (KKT) conditions \cite{CovexBook}, which are detailed for such problems in \cite{zabini,xue}, holds  a strong duality. Therefore, the solution attained in primal form is equal to the dual form. Hence, a near to an optimum resource allocation can be attained while keeping SINR or BER conditions in consideration.

\subsubsection{Round Robin Method} 
The round robin allocation method is formulated in such a way that the same amount of resources are given to all users. In this scenario, a user $j$ perceives the throughput proportional to the one that is possible using all resources \cite{zabini}. Therefore for  a user $j$ , the throughput $\bar T_{ij}$ is given by :

\begin{equation} \label{round robin}
\bar T_{ij} = \frac{\sum_{j\in J}\tilde{T_{j}}}{J}\hspace{0.5cm} \forall j\ \in[1,J]  , 
\end{equation}
 where $J$ is the total number of users in the network and $\tilde{T_{j}}$ is the throughput for the $j^{th}$ user. To simulate such scenarios, we assume that once a connection is established between an eNodeB and its users, the resources are  divided equally among them considering the constraint space of limited SINR.
 
\subsubsection{Opportunistic method} For the  opportunistic case, we formulate the problem as a maximization problem \cite{zabini} in such a manner that each eNodeB allocates maximum RBs to a high throughput link as compared to low throughput links, thus increasing the network throughput non uniformly among users. In the problem formulation,  however, the constraint space will remain the same as the max-min method described earlier. 

\begin{eqnarray}
\max_{i \in I} \hspace{0.2cm}  \arg \left (  \sum_{i=1}^{I}  \left ( \bar{T}_{ij}\times K_{ij}\right) \right ) \hspace{0.5cm}\forall j\ \in [1,J] \nonumber \\
 Y_{ij}\in \{0,1\} \nonumber \\
 \sum_{j=1}^{J}K_{ij} \le N, \hspace{0.5cm} \forall \ i \in [1,I] \nonumber \\
 \sum_{i=1}^{I}Y_{ij} = 1, \hspace{0.5cm} \forall \ j \in [1,J] \nonumber \\
 K_{ij} = Y_{ij}\times N, \hspace{0.5cm} \forall \ i,j \nonumber \\
 K_{ij} \geq (D_{j}/ \bar{T}_{ij})\times Y_{ij}\  \hspace{0.5cm}  \forall \ i,j \nonumber 
\end{eqnarray}

\subsection{Performance Comparisons}

\indent To compare the performance of these methods, we use the very famous  Jain index \cite{jain}, \cite{cheng}. The Jain index for a user $j$ assuming that it  has the throughput  $\tilde{T_{j}}$ is given as:

\begin{equation}
    F= \frac{ \left (\sum_{j\in J}\tilde{T_{j}}\right )^2}  {N \sum_{j\in J}\tilde{T_{j}}^2} \hspace{0.5cm}   \forall  j \ \in [1,J]. \nonumber
   \end{equation}
   
In the next section, we detail out the results obtained by these allocation methods with Jain index as an important performance index.


\section{Simulation results and discussions}
\begin{figure}[!ht]
\centering
\includegraphics[width=8cm,height=6cm]{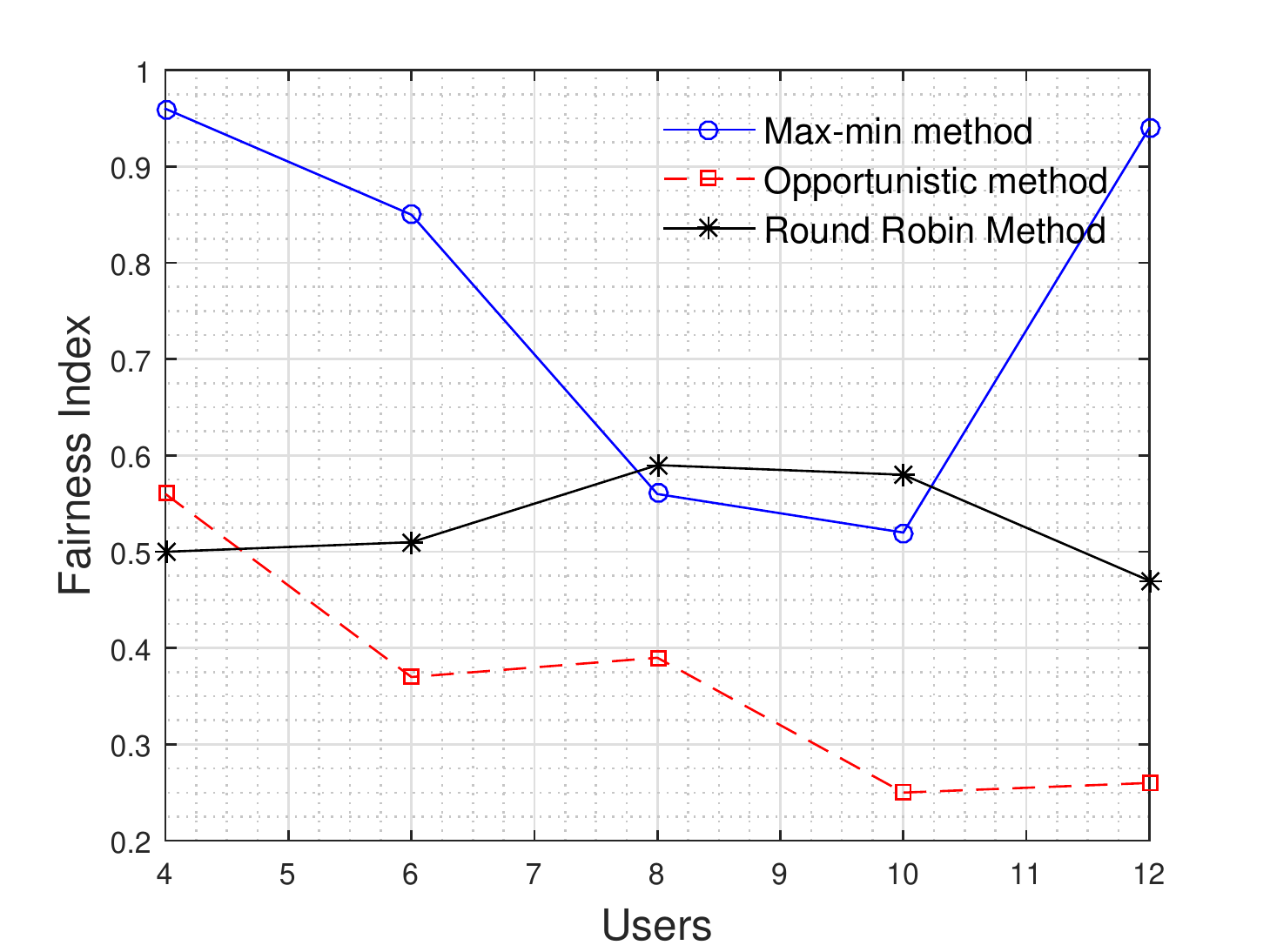}
 \caption{Fairness of algorithms with user density.}
\label{results}
\end{figure}

\begin{figure}[!ht]
\centering
\includegraphics[width=8cm,height=6cm]{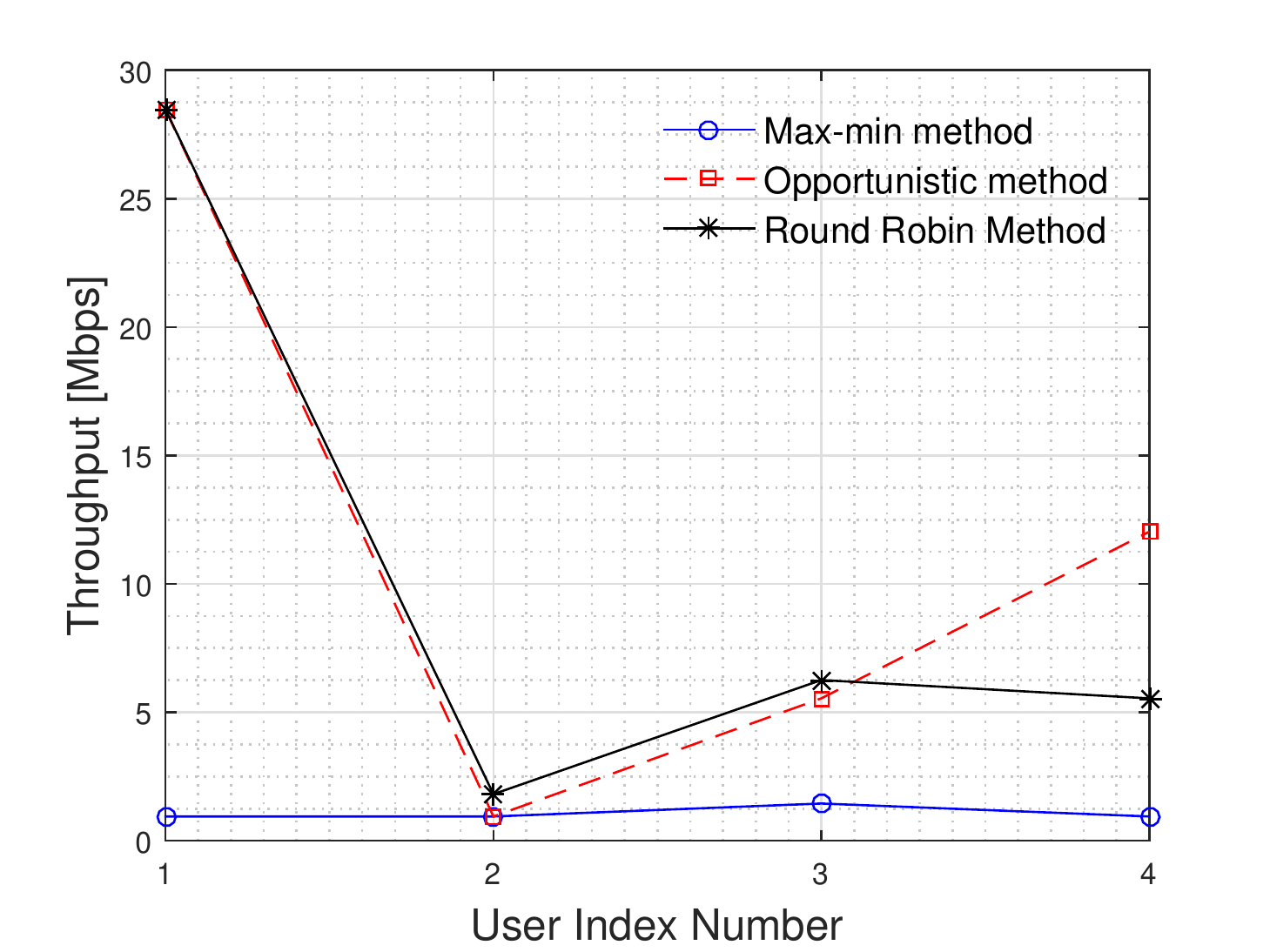}
 \caption{4 user case: Individual user throughput.}
\label{4results}
\end{figure}

\begin{figure}[!ht]
\centering
\includegraphics[width=8cm,height=6cm]{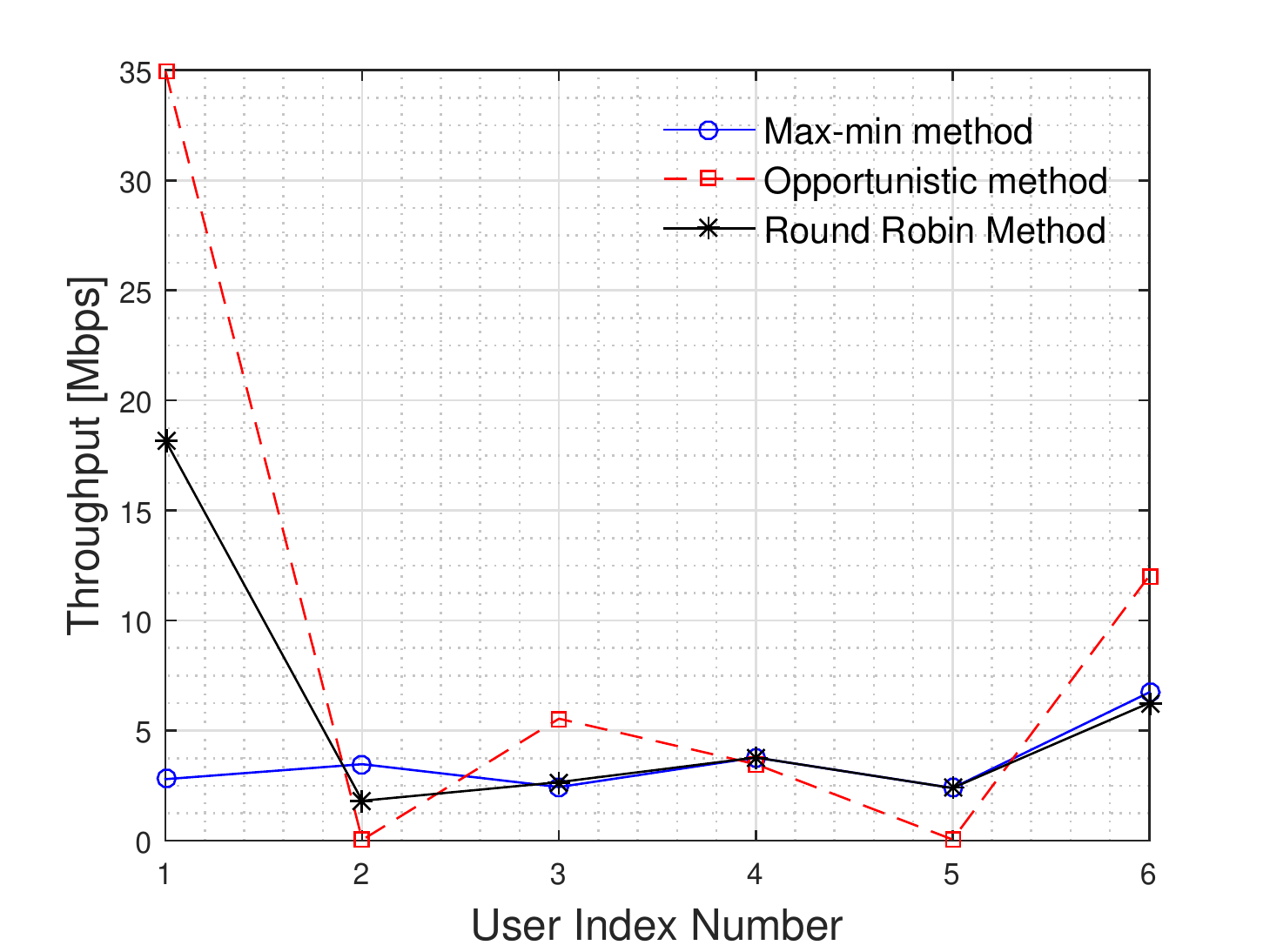}
 \caption{6 user case: Individual user throughput.}
\label{6results}
\end{figure}

\begin{figure}[!ht]
\centering
\includegraphics[width=8cm,height=6cm]{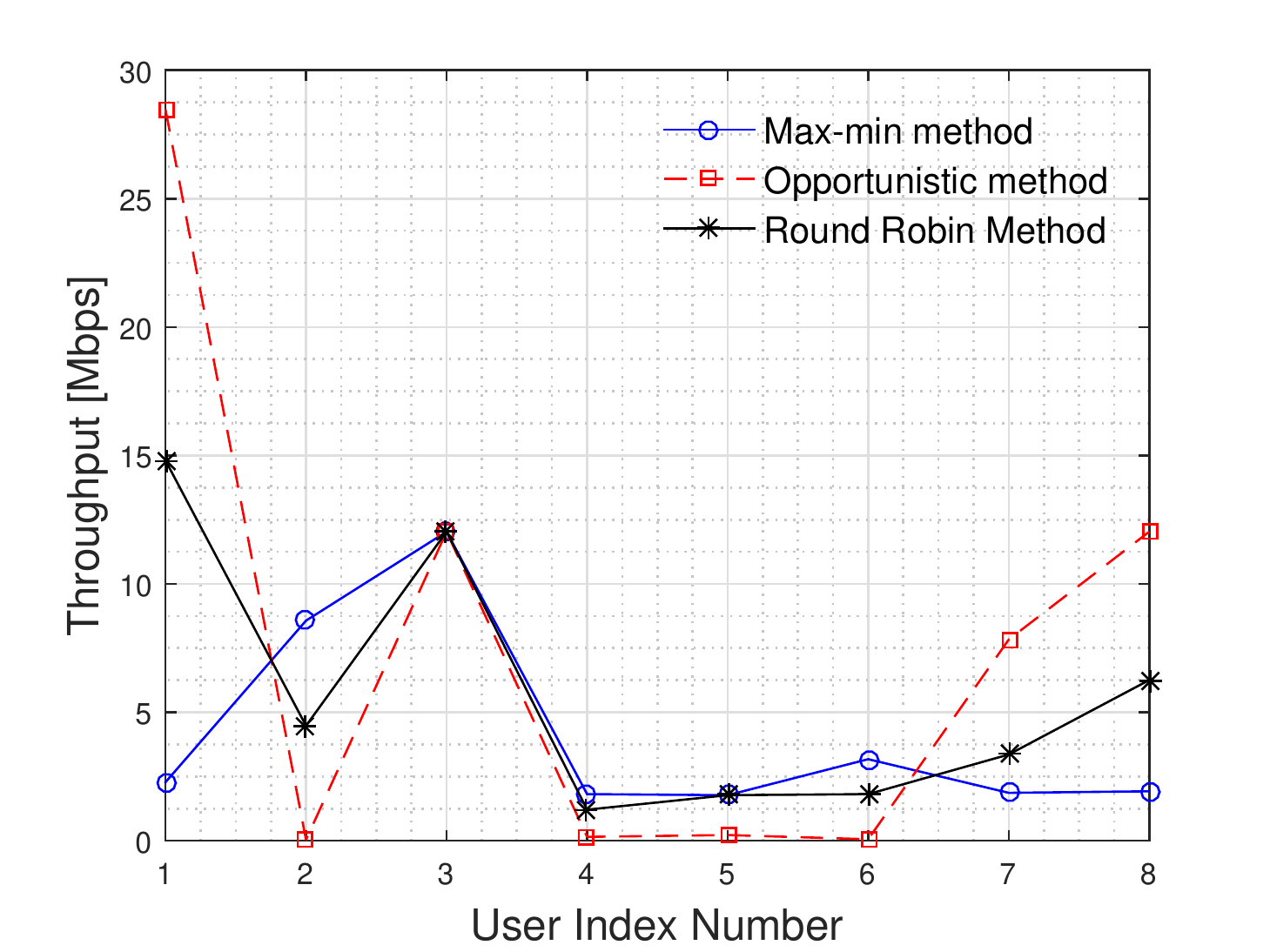}
 \caption{8 user case: Individual user throughput.}
\label{8results}
\end{figure}

\begin{figure}[!ht]
\centering
\includegraphics[width=8cm,height=6cm]{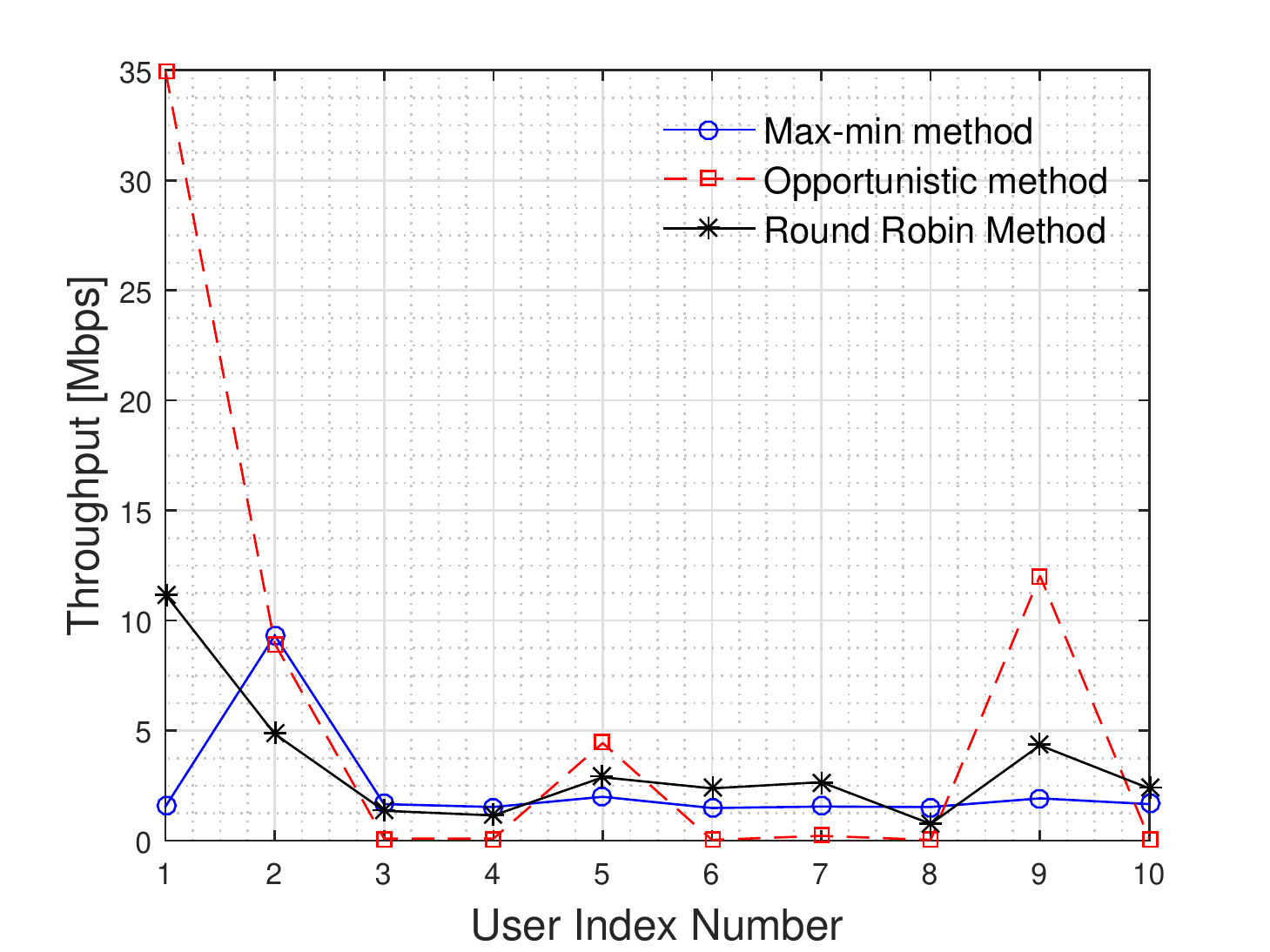}
 \caption{10 user case: Individual user throughput.}
\label{10results}
\end{figure}

\begin{figure}[!ht]
\centering
\includegraphics[width=8cm,height=6cm]{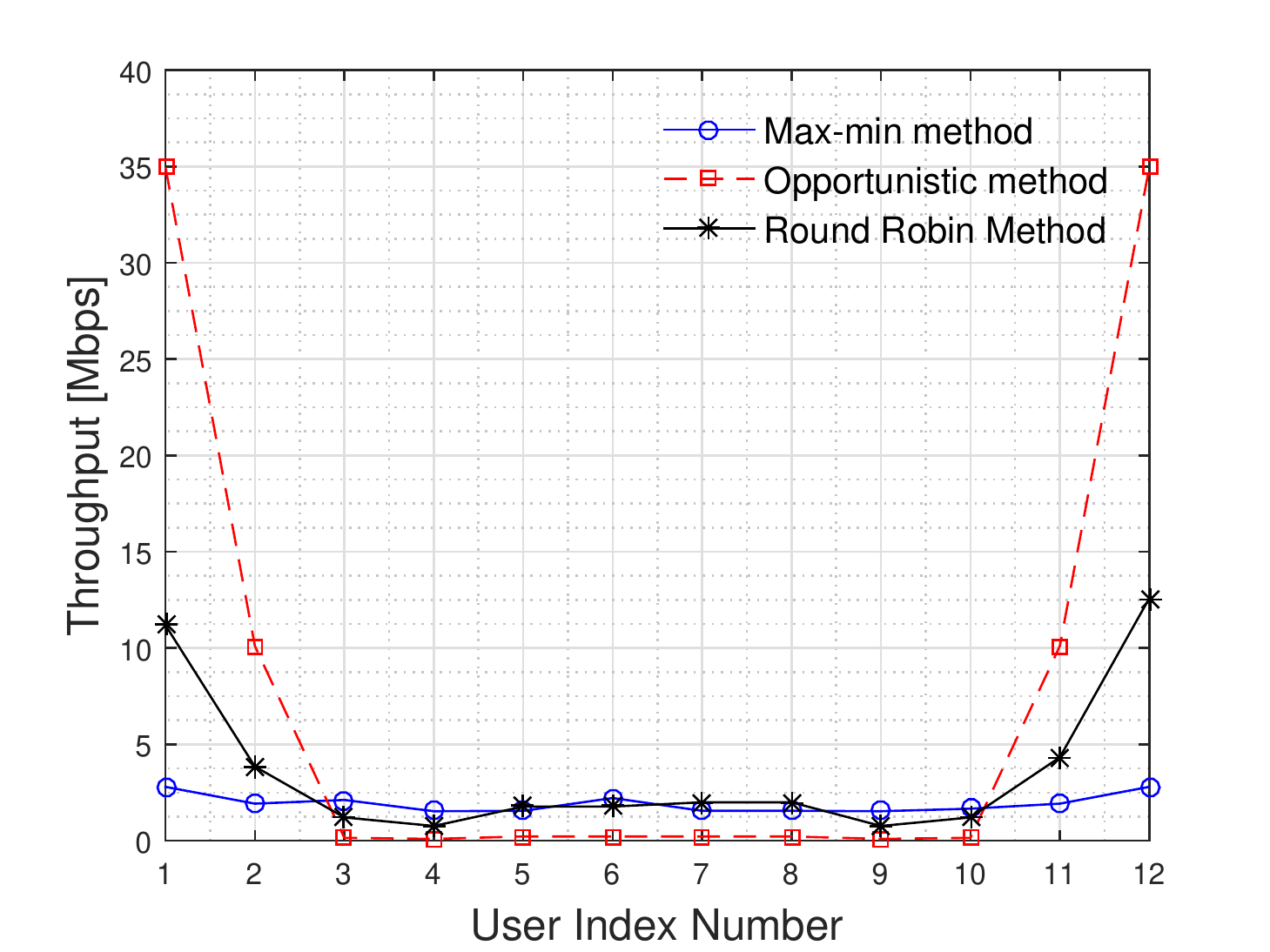}
 \caption{12 user case: Individual user throughput.}
\label{12results}
\end{figure}

In this section, the results obtained by simulating the models described in above sections are compared and discussed. All these  models were simulated in IBM-CPLEX \cite{cplex}. The simulation environment consisted of a sample network scenario in Bosphorus  strait of Istanbul, Turkey (refer to Fig.~\ref{google_map}). The four eNodeBs are land based and are placed on the two opposite ferry ports: Eminonu and Uskudar. The ship positions were assumed to be equidistant to each other in the sea lanes between these two ferry ports and the simulations were carried out with different ship densities in the sea that ranged from 4  to 12 ships.
The resource allocation fairness index (Table~\ref{f}) thus obtained   is plotted in  Fig. \ref{results} and it can be seen clearly  that the  network fairness  with different user densities for max-min method is far better than round robin or opportunistic method.\\
\indent Also, on close analysis of the individual user densities (Figs.~\ref{4results}- \ref{12results}), it can be concluded that the max-min method guarantees better  minimum data throughput per user  as compared to the other two  methods. This is because the max-min method  first allocates RBs to the worst throughput links and then to better throughput links, thus  maximizing the minimum throughput of the links in the network. Moreover, the max-min method balances the overall network throughput uniformly with  increasing number of the user density than the other two optimization methods.

\begin{table}[!ht]
\centering
\caption{Fairness index of different allocation methods with user densities}
\label{f}
\begin{tabular}{|l|l|l|l|l|l|}
\hline
Number of Users & 4 & 6 & 8 & 10 & 12 \\ \hline
Max-Min & 0.96 & 0.85 & 0.56 & 0.52 & 0.94 \\ \hline
Opportunistic & 0.56 & 0.37 & 0.39 & 0.25 & 0.26 \\ \hline
Round Robin & 0.5 & 0.51 & 0.59 & 0.58 & 0.47 \\ \hline
\end{tabular}
\end{table}

\begin{table}[!ht]
\fontsize{9}{9}
\centering
\caption{A sample 8 user resource block allocation}
\label{my-label}
\begin{tabular}{|l|l|l|l|l|l|l|l|l|}
\hline
eNodeB index number  & 2 & 2 & 1 & 4 & 4 & 3 & 4 & 3 \\ \hline
Connected user index number & 1 & 2 & 3 & 4 & 5 & 6 & 7 & 8 \\ \hline
Max-Min Method (RB's) & 2 & 23 & 25 & 12 & 8 & 21 & 5 & 4 \\ \hline
Round Robin  Method (RB's) & 13 & 12 & 25 & 8 & 8 & 12 & 9 & 13 \\ \hline
Opportunistic Method (RB's) & 25 & 1 & 25 & 1 & 1 & 1 & 21 & 25 \\ \hline
\end{tabular}
\label{sample}
\end{table}

\section{Conclusion and future work}

In this  work, we analysed and formulated the   radio resource allocation problem  by  max-min optimization method, which was compared to round robin and opportunistic radio resource block  allocation methods. The scenario involved different ship densities at a time. The max-min integer linear programming method results in the minimum best data throughput that can be guaranteed to  each and every user in a SINR limited sea channels with a high fairness value than the other two methods. The propagation model used for simulation was  3-Ray path loss as compared to very well known 2-Ray path loss  propagation model that is  used extensively for urban landscape. Also, it was observed  that as the number of users begin to increase, the max-min optimization  method performs significantly better with good fairness as compared to the other two  methods. In  future work, we will look into the joint optimization of power and frequency per sub carrier in LTE network over the sea and also would like to extend the work into  the resource allocation problem for 5G networks.

\bibliographystyle{IEEEtran}

\bibliography{ref}

\end{document}